\documentclass[b5paper,twoside]{jpconf}  
\usepackage{graphicx}

\input{psfig}
\def\HI{H{\,\small I}}

\newcommand{\msun}{{$M_\odot$}}
\newcommand{\msunyr}{{$M_\odot$ yr$^{-1}$}}

\begin{document}

\title[Radio jets and outflows of cold gas ]{Radio jets and outflows of cold gas}  

\author[ Raffaella Morganti ]{Raffaella Morganti$^{1,2}$}

\address{$^1$ ASTRON, Postbus 2, 7990 AA Dwingeloo, the Netherlands}
\address{$^2 $ Kapteyn Astronomical Institute, University of Groningen, PO Box 800,
       9700 AV Groningen, the Netherlands}

\ead{morganti@astron.nl} 

\begin{abstract}

Massive gas outflows are considered a key component in the process of galaxy formation and evolution.  It is, therefore, not  surprising that a lot of effort is 
going in quantifying their impact via detailed observations. This short contribution presents recent results obtained from HI and CO observations of different objects where the AGN - and in particular the radio jet - is likely playing  an important role in producing the gas outflows. 
These preliminary results  are reinforcing the conclusion that these outflows have a  complex and multiphase structure where cold gas in different phases (atomic and molecular) is involved and likely represent a major component. These results will also  provide important constraints for establishing  how the interaction between AGN/radio jet and the surrounding ISM occurs and how efficiently the gas should cool to produce the observed properties 
of the outflowing gas. HI  likely represents an intermediate phase in this process, while the molecular gas would be the final stage.   Whether the estimated outflow masses  match what expected from simulations of galaxy formation, it is still far from clear. 

\end{abstract}

\section{A complex interplay}

Understanding galaxy formation and evolution is one of the main challenges for present day astronomy and is one of the key drivers for many of the new and future large telescopes. The most broadly accepted paradigm is that galaxies form by the coalescence of smaller objects, and by the accretion of gas directly from its environment (e.g. \cite{matteo05} and refs. therein). However, a complication that cannot be ignored is the tight interplay between, on the one hand,  star formation/AGN activity and, on the other hand,  the ISM of a galaxy and its surrounding IGM. 
The ISM/IGM provide the material for forming stars and for fueling  the central black hole but, conversely, the enormous energy output from both the stellar and the nuclear activity can have a major influence on the ISM/IGM and  on the further accretion of material onto the galaxy.   Although these feedback effects have been recently emphasized as key ingredient in cosmological simulations of galaxy formation (e.g. quenching star formation), quantifying their impact is still very much a work in progress.

Radio plasma jets are one of the  players in this complicated process and  the fact that the radio-loud phase can be recurrent in the life of a galaxy, makes this component even more important.
Plasma jets can actually have constructive and destructive effects (see \cite{mor10} and refs. therein for an overview). Relevant here is the fact that they can  provide a particularly suitable and fast way of transporting the energy out because they couple efficiently to the ISM/IGM (\cite{tadhunter08}) and inject energy into the large-scale ISM/IGM medium (leaving signatures like X-ray cavities, e.g. \cite{birzan08}) therefore preventing gas to cool to form stars.  However, as we will show below, on galaxy scale is the cold component that appears dominant even in outflows likely originating by jet/ISM interaction. Thus, in the attempt of understanding the effect of  radio-loud AGN in this interplay, more attention has recently been  given to the study (presence, mass and kinematics) of cold gas in the central regions of these objects. 

\section{Background: fast outflows of cold gas} 

The unexpected discover of fast and massive outflows of atomic hydrogen (see e.g. \cite{mor05a} and refs therein) has opened the possibility that the {\sl cold phase of the gas may actually play a role more relevant than expected}. Fast and massive  \HI\  outflows have been found  in young (or restarted) radio galaxies with rich ISM (e.g. bright FIR, molecular gas etc.), see e.g. \cite{mor05a,kanekar08}. In these objects, the radio jet is likely  enshrouded in the medium that has triggered the radio source and is producing outflows while finding its way out. 
In a few cases, the location of the outflow could be identified and found to correspond with off-nucleus regions   {\sl co-located with  bright radio components} (up to a few kpc from the nucleus, see e.g. \cite{mor05b,mor07}). This has further emphasized the role of the radio jet in producing such outflows. 

More recently, outflows  have been detected also in the molecular gas (see e.g. \cite{feruglio10},\cite{ala11}), underlining  even more  the importance of  the cold phase of the gas. The possible relevance of the interaction between radio jets and ISM  is also demonstrated by the strong  H$_2$ lines found in mid-IR spectra  of radio galaxies where jet/cloud interaction is known to occur (\cite{ogle10},\cite{nes11},\cite{guillard12},\cite{papa08}).
Remarkable is the fact that the estimated  mass outflow rates associated with the cold gas can be quite high, ranging from the very extreme case of Mrk231 (\cite{feruglio10}) to the values between  $\sim 1$ and 50 \msunyr  derived from the \HI\ outflows (\cite{mor05a}). Even the latter are  comparable - although at the lower end  - with the mass outflow rates associated with starburst winds (\cite{rupke02}). Interestingly, these values appear to be systematically higher than the mass outflow rates associated with warm gas (\cite{holt11} and ref. therein), while not much is known so far about the contribution from hot gas in these objects.  

Thus, the characteristics and kinematics of the cold gas should be taken into account when estimating the effects of  the nuclear activity. One of the reasons for a growing interest for these outflows is  the fact that they are requested by theoretical models of galaxy formation. However, although gas outflows are now commonly found in AGN, it is not at all clear that their characteristics and impact are consistent with what required by  these cosmological simulation to describe the observed properties of galaxies. 

\begin{figure}
\centerline{\psfig{file=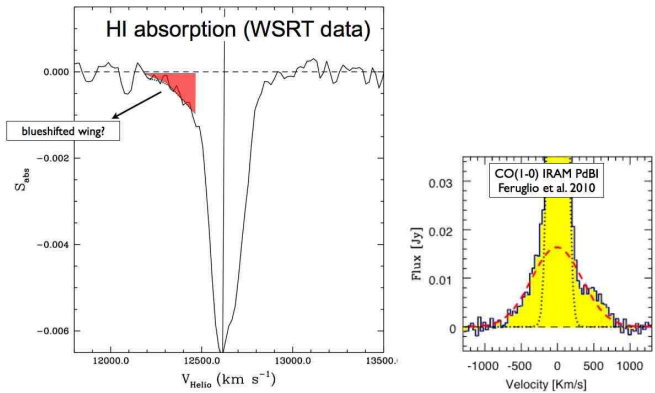,width=9cm,angle=0.}}
\caption{{\sl Left;} \HI\ absorption profile obtained  for Mrk231. The faint blueshifted wing is marked in red. The CO(1-0) profile   is shown on the right side (from \cite{feruglio10}).}
\label{fig:Fig1}
\end{figure}

\section{Importance of multiphase outflows}

Here, we briefly  summarise  recent studies aimed at increasing the number  of objects for which observations are available to trace both the atomic and the molecular components of the outflows. These  studies are strongly limited by the sensitivity of the current instruments but they are  expected to blossom in the coming years  because of  the new  radio facilities coming online.

In the case of the Ultra-Luminous advanced major merger Mrk~231, the recent detection of molecular gas outflow (\cite{feruglio10}) has triggered observations also in \HI\ and the result is shown in Fig. 1. In addition to the already known absorption feature centered on the systemic velocity (and attributed to a circumnuclear disk, \cite{carilli98}), we find a blueshifted wing, albeit  with a smaller amplitude than the CO one. This indicates that the molecular outflow has  associated also a component of neutral atomic gas.  The mass outflow rate derived for the molecular gas is uncertain and can range  from more than 600 \msun yr$^{-1}$ (\cite{feruglio10}) to below $ 100$ \msunyr\ if other assumptions, e.g. the optically thin conversion (see discussion in \cite{ala11}), are used. 
In the first case, the mass outflow rate results much larger than the rate at which gas is converted into stars, therefore suggesting a major impact and an AGN origin for this outflow.   In the second case, the situation would be more uncertain. The outflow rate associated with \HI\ is quite modest (about 10 \msunyr) and clearly lower than the rate associated with the molecular gas. Although Mrk231 is a radio source with a complex structure, the role of the radio plasma (if any) in producing the outflow in this object is not clear and it will have to be investigated with higher resolution observations. 

APEX observations of the radio-loud Seyfert galaxy IC~5063 have also shown a blueshifted wing in the CO(2-1)  profile. The profile suggests that, while  most of the molecular gas is  settle in a large scale disk, a component is  undergoing an outflow similar to what observed in \HI.  IC~5063 was the first object where a fast and massive outflow of \HI\ was detected and located at the position of a bright radio feature (\cite{oosterloo00}).  The mass of the molecular outflow would be about $10^7$\msun\  that can be compared with a few $\times 10^6$ \msun\ associated with the \HI, both much higher than the mass outflow rate of warm, ionised gas (\cite{mor07}). 
Among radio galaxies, it is worth mentioning the case of 3C~305 where new data have revealed that the velocity dispersion of the CO-emitting gas is enhanced at the location of the radio hot spot where the \HI\  outflow is detected (\cite{mor05b}). The CO(1-0) line shows also an asymmetric profile (with a weak blue wing) at this location.  The  analysis of the CO data is now  in progress (Guillard et al. in prep).

Finally, the case of the early-type galaxy NGC~1266 (\cite{ala11}) shows that outflows of cold gas can also occur in  relatively quiet  early-type galaxies.
This object is a  weak radio source ($\sim 10^{21}$ W/Hz) where single-dish and then interferometric observations have found
a massive, centrally concentrated molecular component with a mass of $1.1 \times 10^9$ \msun\ and a molecular outflow with a  mass of $\sim 2.4 \times 10^7$ \msun.    The star formation in NGC~1266 is insufficient to drive the outflow, and thus it is likely driven by the active galactic nucleus. The role of radio plasma in this object is under investigation with higher resolution radio observations.  The estimated mass outflow rate of  $\sim 13$ \msun yr$^{-1}$ leads to a depletion timescale (of the circumnuclear gas)  of $\sim 85$ Myr. This relatively short time may explain why  the phenomenon of gaseous outflow is  rare in early-type  galaxies (\cite{ala11}). 

In summary, these results are reinforcing the idea that AGN- and jet-driven outflows have a {\sl  complex and multiphase structure}  where {\sl cold gas in different phases is involved and likely representing a major component}.  However, despite being massive, these outflows often represent only a very small fraction ($\sim 10^{-4}$) of the Eddington luminosity (see e.g. \cite{mor07}), unlike what required by quasar feedback models (\cite{matteo05}). This is questioning whether the assumptions made in numerical simulations are realistic.
In addition to the implications for feedback effects, the study of outflows is interesting for understanding how the interplay between different components works.  On theoretical grounds  it may actually be expected that most of the mass associated with such outflows is in cold gas and  mostly in the molecular phase  (see e.g. \cite{mellema02}). Two possible scenarios can be envisaged. The gas is entrained and accelerated by the cocoon around the radio jets/lobes  or, alternatively, the  molecular gas is formed out of (efficient) cooling outflowing atomic gas (see also \cite{guillard12}). \HI\ possibly represents an intermediate phase in this cooling process, while the molecular gas would be the final stage.  

\section*{Acknowledgments}
The author  would also like to acknowledge and thank the main  collaborators involved in the projects presented here:  C. Tadhunter, T. Oosterloo, J. Holt, P. Guillard, R. Oonk, W. Frieswijk, K. Alatalo and the ATLAS3D collaboration.

\end{document}